\documentclass[12pt]{article}

\usepackage{amsmath,amsfonts,amssymb,latexsym,stmaryrd,mathrsfs}

\def\be{\begin{equation}}
\def\ee{\end{equation}}
\def\bq{\begin{eqnarray}}
\def\eq{\end{eqnarray}}
\def\beq{\begin{eqnarray*}}
\def\eeq{\end{eqnarray*}}

\begin{document}
\title{\huge{Geodesic incompleteness and  partially covariant gravity}}
\author{\Large{\textsc{Ignatios Antoniadis}$^{1,2}$\thanks{antoniad@lpthe.jussieu.fr}\, and \textsc{Spiros Cotsakis}$^{3,4}$\thanks{skot@aegean.gr}}\\ \\
$^1$Laboratoire de Physique Th\'{e}orique et Hautes Energies - LPTHE\\
Sorbonne Universit\'{e}, CNRS 4 Place Jussieu, 75005 Paris, France\\\\
$^2$ Institute for Theoretical Physics, KU Leuven\\ Celestijnenlaan 200D, B-3001 Leuven, Belgium\\ \\
$^3$Institute of Gravitation and Cosmology, RUDN University\\
ul. Miklukho-Maklaya 6, Moscow 117198, Russia\\ \\
$^4$Research Laboratory of Geometry,\\  Dynamical Systems  and Cosmology,
University of the Aegean\\ Karlovassi 83200, Samos, Greece}
\date{February 2021}
\maketitle
\newpage
\begin{abstract}
\noindent  We study the issue of length renormalization in the context of fully covariant gravity theories as well as non-relativistic ones such as Ho\v{r}ava-Lifshitz gravity. The difference of their symmetry groups implies a relation between the lengths of paths in spacetime in the two types of theory. Provided that certain asymptotic conditions hold, this relation allows us to transfer analytic criteria for the standard spacetime length to be finite  to the Perelman length to be likewise finite, and therefore formulate conditions for geodesic incompleteness in partially covariant theories. We also discuss implications of this result for the issue of singularities in the context of such theories.
\end{abstract}
\newpage
\tableofcontents
\newpage
\section{Introduction}
Non-relativistic theories of gravity have been very popular recently for a number of different reasons. In the direction of renormalizing general relativity,  a prime example is provided by the Ho\v{r}ava-Lifshitz theory where a power-counting renormalizable theory is offered by the drastic abandoning of Lorentz symmetry and introducing an anisotropic scaling between space and time \cite{h1,h2}, \cite{ba,b2}. This has major implications for a number of areas, especially in cosmology, where general relativity is partially abandoned and new difficulties arise due to the presence of higher-order terms and the idea of scalar graviton, but also new light is shed to old issues like the possibility of having a scale invariant spectrum of cosmological perturbations even without inflation, a bouncing cosmology, or a possible non-chaotic evolution towards the singularity of a Bianchi IX cosmology \cite{bblp}-\cite{amm}.

Due to the close relation of the Ho\v{r}ava-Lifshitz equations to a generalized form of the Ricci flow equation, new ways are possible for the construction of topological quantum gravity theories by utilizing special mathematical properties of the later \cite{h3}. Secondly, the Ricci flow as well as other non-relativistic geometric flows such as the mean-curvature flow have long been known for their interesting properties especially with regard to the formation of singularities \cite{ric1}-\cite{ck}, in association with corresponding properties known for the Einstein flow \cite{ycb}-\cite{mon}, although a clear relation between the nature of singularities between different non-relativistic flows and general relativity is presently elusive. Thirdly, the importance of alternative geometries such as the Newton-Cartan (NC) geometry in gravitation through a non-relativistic expansion is useful in delineating more precisely  the relations of non-relativistic gravity and general relativity \cite{ho1}-\cite{ho2}.

A distinctive feature common to all the non-relativistic theories above is that their symmetry, or \emph{invariance}, groups are not the full diffeomorphism group (as in the case of a theory such as general relativity) but typically some proper subgroup of it. This results in a different behaviour of the solutions (i.e., the space-time metric and various fields) of a non-relativistic theory as opposed to those of general relativity, in particular  with respect to their transformations by elements of their symmetry group (cf. \cite{an}, \cite{strau}. The solutions of the theory are of course invariant under the symmetry group, and constitute the theory's \emph{absolute elements}. In general relativity, and in distinction with all the non-relativistic theories mentioned above, there are no absolute elements because the symmetry group coincides with the covariance group (GR is `generally covariant'). In a non-relativistic theory, we may of course write down a metric-solution of the field equations of the theory-which looks the same as  in general relativity, however,  the subtle difference in their symmetry groups mentioned above will reflect important changes in the causality properties associated to the two metric structures.

The purpose of this note is to study this point. In the next Section, we introduce this difficulty and in Section 3 we suggest a new approach for the study of causality in the two frameworks of a fully and a partially covariant theory. In Section 4, we establish the existence of an asymptotic relation between the length functional for paths in space-time typically used in a non-relativistic framework and the standard spacetime length. This relation will allow us in Section 5 to find  conditions for partially covariant flows to be geodesically  incomplete from those for the spacetime length in fully covariant theories. We discuss our results in the last Section.

\section{Ho\v{r}ava-Lifshitz vs. spacetime geometry}
In this Section,  we compare the Ho\v{r}ava-Lifshitz theory to general relativity, especially their causal structures, focusing on their covariance groups. This is a necessary first step since we shall import ideas and methods from general relativity to the framework of the former theory. As discussed in the Introduction, what we describe below is valid not only for the  Ho\v{r}ava-Lifshitz theory but also for any other non-relativistic theory based on NC geometry, a well as on other geometric flows such as the Ricci flow.
%
%

It is well known that in general relativity there is a way,  pioneered by Yvonne Choquet-Bruhat,  to split the 4-metric, connection, and curvatures, in a Cauchy-adapted frame (cf. \cite{ycb} and refs. therein). Such a decomposition is very suitable for the study of the initial value formulation of the theory (as well as its Hamiltonian  (ADM) formulation). In this case, the evolution equations are expressed in terms of the spatial 3-metric  $g_{ij}$ and its extrinsic curvature $K_{ij}$, but there are further \emph{gauge variables}, the lapse $N$ and the shift $N_i$. These,  together with the spatial metric,  make up the 4-metric $(g_{ij},N,N_i)$, satisfying the Einstein equations and being invariant under the full diffeomorphism group, $\textrm{Diff}(\mathcal{V})$, of the spacetime manifold $\mathcal{V}$.

But in the Ho\v{r}ava-Lifshitz theory, because of the restricted covariance of the field equations under (the subgroup of  $\textrm{Diff}(\mathcal{V})$) $\textrm{FPDiff}(\mathcal{V})$ of foliation-preserving diffeomorphisms, the functions $N,N_i$ are no longer gauge variables but are now promoted to \emph{fields} on the 4-manifold $\mathcal{V}$. The `4-metric' $(g_{ij},N,N_i)$, solution of the Ho\v{r}ava-Lifshitz  field equations, has now a smaller invariance group than in general relativity. This has major implications for the causal structure of the theory. Below we briefly discuss some of these implications in the context of the  Ho\v{r}ava-Lifshitz theory.

On the 4-manifold $\mathcal{V}=\mathcal{M}_{t}\times [t_{0},\infty ], t_0\in\mathbb{R}$, we are given the following data: a smooth Riemannian metric $g_{ij}$ on the 3-manifold $\mathcal{M}_{t}=\mathcal{M}\times\{t\}$, a smooth function $N(t,x^i)$ defined on $\mathcal{V}$, and a vector field $N^i(t,x^j)$ tangent to the 3-manifold $\mathcal{M}$. As we shall show below, the data $N,N^i$, can play a role analogous to that of the lapse function and shift vector of the usual $(3+1)$-formalism in general relativity. A basic geometric assumption of the Ho\v{r}ava-Lifshitz gravity  is the existence of a `book-keeping' line element form,
\begin{equation}\label{hlmetric}
ds^2=-N^{2}dt^{2}+g_{ij}(t,x)\left(dx^{i}+N^{i}dt\right)\left( dx^{j}+N^{j}dt\right).
\end{equation}

It is usually assumed that $N$ is  a function of $t$ only (the so-called `projectable' case), but most importantly the form (\ref{hlmetric}) considered as a solution of the Ho\v{r}ava-Lifshitz field equations (identical to the Ricci flow equations when $N=1,N_i=0$),
\be\label{ricflow}
\dot{g}_{ij}=-\frac{\kappa^2}{2\kappa_{W}^2}N\left(R_{ij}-\frac{2\lambda -1}{2(D\lambda-1)}g_{ij}R \right)+\nabla_{i}N_{j}+\nabla_{j}N_{i},
\ee
is invariant under the action of the restricted group of foliation-preserving diffeomorphisms $t\rightarrow\tilde{t}(t),x\rightarrow\tilde{x}(t,x)$, not of the full group of spacetime transformations \cite{h1}, \cite{h2}, \cite{ba}, \cite{bblp}. Despite the identical form of $ds^2$ above to the standard `adapted-frame' space-time decomposition of a metric as in general relativity (cf. e.g., \cite{ycb}),   (\ref{hlmetric}) is not the usual proper time,  and equipping the 4-manifold $\mathcal{V}$ with the form (\ref{hlmetric}) does not make $(\mathcal{V},g_{HL})$  a spacetime in the general relativistic sense of the word.

In particular, there is no null  structure on $(\mathcal{V},g_{HL})$ like in general relativity, nor a relativistic notion of causality or chronology, nor can there be the usual trichotomy of timelike, null, spacelike, for vectors at any point $p$ on $\mathcal{V}$. These  `peculiarities' are also shared by other non-relativistic theories of  gravity such as Newton-Cartan theories, as well as geometric flows such as the Ricci flow, cf. \cite{ho1}, and \cite{ric1}, Section 14.1, respectively.  In fact, in such theories, the classical metric (\ref{hlmetric}) is just the leading term in an $1/c$ expansion about zero ($c\rightarrow\infty$) \cite{h1,h2,ho2}. For a given curve $C:I\subset\mathbb{R}\rightarrow \mathcal{V}$ on the manifold $\mathcal{V}$, there can be no invariant definition of a notion of proper length for $C$ using the form (\ref{hlmetric}) in such an approximation, no notion of causal geodesics on $(\mathcal{V},g_{HL})$, hence no obvious way to speak of the usual route through geodesic (in-)completeness to spacetime singularities, maximal curves, global hyperbolicity, etc.

Hence, although there is an obvious interest in the further analysis and examination of the implications and classical meaning of non-relativistic theories with anisotropic scaling which also share desirable quantum properties \cite{h1,h2}, given their restricted invariance (of the line element $ds^2$ as in Eq. (\ref{hlmetric}) and action) under only foliation-preserving diffeomorphisms, there is an apparent  difficulty to talk sensibly about dynamics and asymptotic properties of spacetime fields near singularities in terms of standard spacetime notions. How are we to somehow import a generic notion of geodesic (in-)completeness into these frameworks and therefore be able to compare  non-relativistic gravity theories to the more usual ones that allow a full spacetime interpretation?

\section{A new approach}
Consider two points $p,q\in\mathcal{V}$ belonging to a suitable, globally hyperbolic region $\mathcal{N}$ of $\mathcal{V}$ with $q$ in the future of $p$ in the spacetime metric $g_4$, and introduce Minkowski normal coordinates $(t,x^i)$ in the region  $\mathcal{N}$ (that is $\partial _t$ is timelike and future-pointing and the null cone $T_p\mathcal{V}$ is the set $t^2-\sum (x^i)^2=0$). We can follow standard arguments (cf. e.g., \cite{pen72}, Proposition 7.2, \cite{he73}, Lemma 6.7.2) and introduce Gaussian normal coordinates $T,X,Y,Z$ on $\mathcal{V}$, where $T=(t^2-\sum (x^i)^2)^{1/2}$, $X^1=x^1/t,X^2=x^2/t,X^3=x^3/t$. This is the so-called \emph{synchronous system} in which the surfaces $T=\textrm{const.}$ are spacelike while the curves $X^i=\textrm{const.}$ are timelike geodesics orthogonal to these. The metric $g_4$ then takes the standard form,
\be\label{synch}
ds^2_{GR}=dT^2-g_{ij}dX^idX^j.
\ee
Setting $p=C(T_0),q=C(T_1)$,  the spacetime length of any curve $C$ connecting the points $p,q\in\mathcal{N}$ is given by (a dot denotes differentiation with respect to $T$),
\be\label{g-l}
l_{GR}(C)=\frac{1}{T_1-T_0}\int_{T_{0}}^{T_{1}}\left(1-g_{ij}\dot{X}^i\dot{X}^j\right)^{1/2}dT,
\ee
and so the length is equal to 1 for the curves $X^i=\textrm{const.}$ (the 3-metric $g_{ij}$ is positive-definite). That is, the geodesic connecting the two points $p,q$ has maximum length.

We note that this  process of obtaining maximal geodesic curves described above is not available in the Ho\v{r}ava-Lifshitz theory (at least in its standard `projectable' version where $N=N(t)$). This is simply because it is not feasible to pass to synchronous coordinates which have $N$ depending on both $t$ \emph{and} the spatial coordinates $x^i$ as above (even a choice of coordinates which translates  $ds_{HL}^2$ to the form  (\ref{synch}) would not do, because the curves $X^i=\textrm{const.}$ are not  geodesics  on $\mathcal{V}$).


However, we may proceed as follows. Given the  resemblance of the Ho\v{r}ava-Lifshitz field equation to the generalized Ricci flow (\ref{ricflow}), it is not unreasonable to think of using the $\mathcal{L}$-length function (the so-called Perelman length) as a possible means of addressing geodesic problems (dependent on its second variation) in the framework of Ho\v{r}ava-Lifshitz and other non-relativistic theories. We show below that the $\mathcal{L}$-length function is in fact asymptotically connected to the standard spacetime length of fully covariant theories. One way to show this is to use a similar  procedure to the length renormalization method for the so-called potentially infinite metrics. It is known that for the Ricci and other Riemannian curvature flows, this method can be used to derive the Perelman $\mathcal{L}$-length (cf. e.g., Chapter 11 of \cite{cln}).

\section{Renormalization of the spacetime length}
What we need is a procedure that applies equally  to both non-relativistic and fully covariant metric theories.
For this purpose, we introduce now the  family of metrics $g_{\xi}$ which have  the same  generic form of both the fully covariant metric $g_4$ used earlier to discuss the spacetime length $l_{GR}(C)$, as well as the foliation-preserving-diffeomorphisms-invariant metric $g_{HL}$ used earlier for the non-relativistic length function $l_{HL}(C)$.  Following \cite{per1}, for the family of  metrics $g_{\xi}$  we choose the lapse and shift functions such that,
\be\label{lapse}
g_{\xi}:ds_{\xi}^2=-N_{\xi}^{2}dt^{2}+g_{ij}(t,x)dx^{i}dx^{j},\quad -N_{\xi}^2(t,x^i)=R(g_{ij})(t,x^i)+\frac{\xi}{2t}<0,\quad N_{\xi}^i=0,
\ee
with $R$ being the scalar curvature of the 3-metric $g_{ij}(t,x^i)$, and $\xi$ a  real negative constant large enough such that the above inequality for the scalar curvature is true.
%

Then the standard spacetime  length of any curve $C:(t_0,t_1)\rightarrow\mathcal{V}$ with respect to the metrics $g_{\xi}$ is given by,
\be\label{ycbl}
l_{\xi}(C)=\int_{t_{0}}^{t_1}\left(-N_{\xi}^2+g_{ij}\dot{C}^i\dot{C}^j\right)^{1/2}dt
=\int_{t_{0}}^{t_1}\left(R+\frac{\xi}{2t}+\left|\frac{dC}{dt}\right|^2_{g(t)}\right)^{1/2}dt,
\ee
and this is well-defined provided the $g_{ij}$-length of $C$ is bounded from below. Then we write the integrand as
\be
\left(\frac{|\xi|}{2t}\right)^{1/2}\left(1+x \right)^{1/2},\quad x=\frac{2t}{|\xi|}\left(R+\left|\frac{dC}{dt}\right|^2_{g(t)}\right),
\ee
and expand $(1+x)^{1/2}$ keeping only up to the highest order non-trivial term in $\xi$. We find
\be\label{asym}
l_{\xi}(C)=\int_{t_{0}}^{t_1}\left(\left(\frac{|\xi|}{2t}\right)^{1/2}+\frac{|\xi|^{-1/2}}{\sqrt{2}}\sqrt{t}
\left(R+\left|\frac{dC}{dt}\right|^2_{g(t)}\right)
+O(|\xi|^{-3/2})\right)dt,\quad|\xi|>1.
\ee
In this formula, of course, there are conditions - see below - for the spacetime length $l_{\xi}(C)$ to be related to an \emph{extracted length} corresponding to the second term in the integral, that is the  Perelman length function $l_{per}(C)$ for the spacetime curve $C$, given by,
\be\label{per}
l_{per}(C)=\int_{t_{0}}^{t_{1}}\sqrt{t}\left(R+\left|\frac{dC}{dt}\right|^2_{g(t)}\right)dt,
\ee
(the so called reduced length is $l_{per}(C)/2(\sqrt{t_1}-\sqrt{t_0})$, cf. \cite{per1}). We recall that in distinction to the spacetime length which maximizes the length (proper time) between two points in spacetime, the Perelman length is a minimum along geodesics (cf. (\cite{ric1})-(\cite{per1})). However, since the first integral in the right-hand-side of  (\ref{asym}) is finite, it follows that the spacetime length $l_{\xi}(C)$ is finite if and only if the Perelman length $l_{per}(C)$  is finite. The conditions under which this is possible are given in the next Section.




The fact that the Perelman length (\ref{per}) may be extracted from Eq. (\ref{asym}) and so be connected with the usual spacetime length $l_{\xi}(C)$ asymptotically for the family of metrics $g_{\xi}$ is very important. It shows that there may be a connection between the asymptotic nature of the fields near  singularities met in various  non-relativistic geometric flows such as the Ricci flow \cite{ha}, \cite{ck}, and that near the `physical' spacetime singularities  defined as geodesic incompleteness in general relativity. This is in fact the case, as we demonstrate  in the next section.

\section{Completeness and singularities}\label{comp}
It is well known  that spacetime singularities and finite geodesic length as well as geodesic completeness and infinite such length may be described, not in the usual 4-dimensional sense of the singularity theorems in general relativity, but through analytic criteria based on conditions imposed on quantities appearing in the $(3+1)$-decomposition of spacetime, such as the extrinsic curvature of the time slices, \cite{ycb}, \cite{ycb1}. 
As we show below, this approach is particularly useful in discussing geodesic (in-)completeness in non-relativistic frameworks as well.

There are two main results of this type giving sufficient conditions for causal geodesic completeness as well as of incompleteness, for the time interval of the form $[t_0,\infty)$, cf. \cite{ycb}, \cite{ycb1}. Because of the result we proved in the previous Section, we may examine whether conditions proved for the spacetime length in Refs. \cite{ycb}, \cite{ycb1} may be true for the Perelman length and non-relativistic (partially covariant) frameworks. Our time intervals presently are necessarily finite of the form $[t_0,t_1]$,  \emph{not} infinite like in the Refs. \cite{ycb}, \cite{ycb1}. This is not a restriction, however. The simple choice of a new time parameter $\tau$, given by
\be
\tau=\frac{1}{t_1-t},
\ee
will transform the spacetime length integrals of the form $\int_{t_0}^{t_1}f(t)dt$ considered here in  Eq. (\ref{asym}), to the length integral,
\be
\int_{\frac{1}{t_1-t_0}}^{\infty}g(\tau)d\tau ,\quad g(\tau)=f\left(t_1-\frac{1}{\tau}\right)\frac{1}{\tau^2},
\ee
and hence, the conditions found in \cite{ycb}, \cite{ycb1} will apply in the present context unaltered. Here we assume that $f$ is continuous and unbounded on $[t_0,t_1)$, but an analogous change of time variable, $\tau=\frac{1}{t-t_0}$, will transform an improper integral on $(t_0,t_1]$ to one on $[1/(t_1-t_0),\infty)$ for the function $h(\tau)=f\left(t_0+\frac{1}{\tau}\right)\frac{1}{\tau^2}$.

For causal geodesic completeness (that is, infinite geodesic length) the conditions may be phrased as follows (cf. \cite{ycb1} for definitions and complete proofs). Suppose that the spacetime  $(\mathcal{V},g_4)$ with  metric given by (\ref{hlmetric}) is such that the data $N,N^i,g_{ij}$ are all uniformly bounded. Then if the spatial norms of the space gradients of the lapse, $(\nabla_i N)^2$, and of the extrinsic curvature, $K_{ij}K^{ij}$, are both bounded by time-dependent functions (which are integrable on $(t_0,\infty)$), then $(\mathcal{V},g_4)$ is timelike and null future geodesically complete. Equivalently, we may rephrase these conditions on the extrinsic curvature using its traceless part $P_{ij}=K_{ij}-(1/3)g_{ij}K$  \cite{ycb1}. If the spatial norm of $P$,
\be
|P|_{g_{ij}}=K_{ij}k^{ij}-\frac{1}{3}K^2,
\ee
is bounded on $(t_0,\infty)$, then the spacetime  $(\mathcal{V},g_4)$ is future timelike and null g-complete.
(We may also rephrase these sufficient conditions for completeness as \emph{necessary} ones for \emph{in}completeness, that is when $(\mathcal{V},g_4)$ is singular.  Then $K_{ij}K^{ij}$ must blow up (assuming that the gradient of the lapse is bounded). Similarly, assuming that we have a singular spacetime, then we may use the traceless part to say that $P_{ij}$ must blow up somewhere.)

Now we apply the above completeness results to the length function $l_{\xi}(C)$ given by the first equality of Eq. (\ref{ycbl}), Under the same sufficient  conditions the metric $g_{\xi}$ given by  (\ref{lapse}) will also be future geodesically complete. Namely, for an infinite length $l_{\xi}(C)$ (that is $g_{\xi}$ is future geodesically complete), we require that $(\nabla_i N_\xi)^2$ and $K_{\xi,ij}K_\xi^{ij}$
are uniformly bounded. However, if the length $l_{\xi}(C)$ given in Eq. (\ref{asym}) diverges (as it is necessary for completeness), we cannot separate the integral and extract the Perelman length (\ref{per}) even though the integral of the first term converges on $[t_0,t_1]$. Therefore starting from an infinite $l_{\xi}(C)$ length and looking for sufficient conditions for completeness is not useful for our purposes, as the situation becomes inconclusive.

However, we may suppose that the length $l_{\xi}(C)$ is finite (i.e., geodesic incompleteness), and look for \emph{necessary} conditions for incompleteness. This situation is explained in great detail in Refs. \cite{iklaoud1,iklaoud2}, and there are various cases to consider according to which some of the quantities $(\nabla_i N_\xi)^2$ or $K_{\xi,ij}K_\xi^{ij}$ blow up or become discontinuous in some way. These will provide new criteria for the Perelman length to be finite. Therefore  geodesic incompleteness is guaranteed in both frameworks, Ricci flow and general relativity.


Further, we look for \emph{sufficient} conditions for incompleteness: These  are that $(\nabla_i N_\xi)^2$ is integrable and that $K_{\xi,ij}v^i v^j\geq k>0$, for all spatial vectors $v$, cf. Ref. \cite{ycb}. These conditions will also give sufficient criteria for the Perelman length to be finite in this context.

\section{Discussion}
In this note we have considered possible relations between fully covariant theories and non-relativistic ones, especially with regard to their causality structures and the issue of singularities (in the sense of geodesic incompleteness). Since two such theories are very different due to the existence of `absolute elements', it is not obvious how their causal structures could be related if at all. This is also an issue that may arise  when one asks in a non-relativistic context whether or not  some property deduced from a  special solution has possibly a more general (i.e., `generic') meaning or significance.

The main result we reported in this work is that if the spacetime length  of paths is finite, then the Perelman length is also finite asymptotically if certain conditions hold. This suggests a possible relation between the spacetime singularities of general relativity associated with geodesic incompleteness, and those of non-relativistic flows such as the Ricci flow, the Ho\v{r}ava-Lifshitz theory, or in the context of Newton-Cartan geometry. We believe that this relation extends without much change in the setting of any diffeomorphism invariant theory vs. a non-relativistic one.

It is therefore possible that, with certain adjustments, the proofs of key results in global causality theory in general relativity may be  applicable in the wider context of non-relativistic gravity. However, the exact formulation of something like this is at present totally unknown. Conversely, some techniques in Ricci flow (also applicable in principle to other partially covariant flows) may become useful in situations where the spacetime length is finite and a possible extension of the solution is necessary. Unfortunately, our present results cannot deduce an infinite Perelman length from geodesic completeness (infinite spacetime length). However, it follows from our incompleteness result about the finiteness of the spacetime length,  that if we know that the Perelman length is infinite, then the spacetime length will also be infinite. Therefore geodesic completeness in a partially covariant theory would imply standard spacetime geodesic completeness under the other conditions considered in this paper. In the Ricci flow, an infinite Perelman length is obtained by the process of surgery at specific times, the singular times. This raises the intriguing question whether one could similarly continue the Einstein flow using surgery at the singularities in general relativity. We hope to return to this problem in the future.

\section*{Acknowledgments}
Work partially performed by I.A. as International professor of the Francqui Foundation, Belgium.
This paper is dedicated to the fond memory of our beloved friend and colleague, the late Ioannis Bakas. Giannis was not only a brilliant mathematical physicist, but also a extremely kind, generous, pleasant, and honest person. He had a very wide knowledge and deep interest in both theoretical physics and pure mathematics, as this is clearly testified by his scientific work. He took a very active part in early discussions of this project, and had he lived to this day, this paper would have been a better one.

\end{document}